\documentclass[twocolumn,amsmath,amssymb,superscriptaddress]{revtex4}
\usepackage{graphicx}
\usepackage{color}
\usepackage{url}
\usepackage{gensymb}
\usepackage{epsf}

\begin{document}

\title{Direct evidence for competition between the pseudogap and high
temperature superconductivity in the cuprates}
\author{Takeshi Kondo}
\affiliation{Ames Laboratory and Department of Physics and Astronomy, Iowa State
University, Ames, IA 50011, USA}
\author{Rustem Khasanov}
\affiliation{Laboratory for Muon Spin Spectroscopy, Paul Scherrer Institut, CH-5232
Villigen PSI, Switzerland}
\author{Tsunehiro Takeuchi}
\affiliation{Department of Crystalline Materials Science, Nagoya University, Nagoya
464-8603, Japan}
\affiliation{EcoTopia Science Institute, Nagoya University, Nagoya 464-8603, Japan}
\author{Joerg Schmalian}
\affiliation{Ames Laboratory and Department of Physics and Astronomy, Iowa State
University, Ames, IA 50011, USA}
\author{Adam Kaminski}
\affiliation{Ames Laboratory and Department of Physics and Astronomy, Iowa State
University, Ames, IA 50011, USA}
\date{\today }

\maketitle

{\bf
A pairing gap and coherence are the two hallmarks of superconductivity.
In a classical BCS superconductor they are established simultaneously
at $T_{\rm c}$. In the cuprates, however, an energy gap (pseudogap) extends
above $T_{\rm c}$ \cite{Emery, Timusk, HongPseudogap, LoeserPseudogap, NormanNature, DamascelliReview, CampuzanoReview, Anderson}. The origin of this
gap is one of the central issues in high temperature superconductivity.
Recent experimental evidence demonstrates that the pseudogap and the
superconducting gap are associated with different energy
scales\cite{Raman, ShenScience, Twogap, ShenNature, EricTwogap,
HongCondMat}. It is however not clear whether they coexist
independently or compete\cite{Raman, Twogap, EricTwogap, Rustem}. In
order to understand the physics of cuprates and improve their
superconducting properties it is vital to determine whether the
pseudogap is friend or foe of high temperature supercondctivity \cite{Norman05}. Here we report evidence from angle resolved photoemission
spectroscopy (ARPES) that the pseudogap and high temperature
superconductivity represent two competing orders. We find that there is
a direct correlation between a loss in the low energy spectral weight
due to the pseudogap and a decrease of the coherent fraction of paired
electrons. Therefore, the pseudogap competes with the superconductivity
by depleting the spectral weight available for pairing in the region of
momentum space where the superconducting gap is largest. This leads to
a very unusual state in the underdoped cuprates, where only part of the Fermi surface develops coherence.}

Coherence in the superconducting state of the cuprates manifests itself by
the appearance of a narrow peak in the ARPES lineshape \cite{Fedorov}, while
the pseudogap \cite {HongPseudogap,LoeserPseudogap,NormanNature,Twogap} depletes the low energy spectral weight below the pseudogap energy. The simplicity of the Bi$_2$Sr$_2$CuO$_{6+\delta}$ (Bi2201) spectra, as measured by ARPES, permits us to perform a straight forward quantitative analysis of the two features because the energy distribution curves (EDCs) in this single layer material lack the large renormalization effects (e.g. peak-hump-dip structure) and bilayer splitting that are present\cite{DamascelliReview, CampuzanoReview} in double layered Bi$_2$Sr$_2$CaCu$_2$O$_{8+\delta}$ (Bi2212). This feature, however, means the spectral changes associated with the superconducting transition in Bi2201 are much more difficult to observe\cite{FengBi2201}. By acquiring very high resolution and stable ARPES data with high statistics, we are able to study the temperature and momentum dependence of the spectral weight near the chemical potential, with unprecedented accuracy. Experimental and sample preparation details are provided in the Supplementary Information. In Fig. 1 we examine the temperature dependence of the spectral lineshape in overdoped Bi2201 ($T_{\rm c}$=29K). Above the pseudogap temperature ($T^{\ast }$) ($\sim$110K for this sample), the symmetrized EDCs \cite{NormanNature} (see Supplementary Information) show a peak centered at the chemical potential - consistent with the metallic state of the sample. Upon cooling below $T^{\ast }$, the low energy spectral weight decreases (within $\sim $20 meV), leading to a characteristic dip and very broad spectral peaks that signify the opening of an energy gap, as shown in Fig. 1(d). The loss of the low energy spectral weight continues all the way to $T_{\rm c}$ (Fig. 1(e)). As the temperature is decreased below $T_{\rm c}$ a small but very sharp peak with a width of $\sim $10meV appears at a binding energy of $\sim $15 meV. The intensity of this peak increases with decreasing temperature. The presence of a coherent peak in the superconducting state is also evident in the ARPES intensity maps shown in Fig. 1(b)-(c), where a thin
line of higher intensity appears at $\sim $15 meV below $T_{\rm c}$. It is quite
remarkable that the presence of this peak is closely associated with the
\begin{figure}
\includegraphics[width=3.6in]{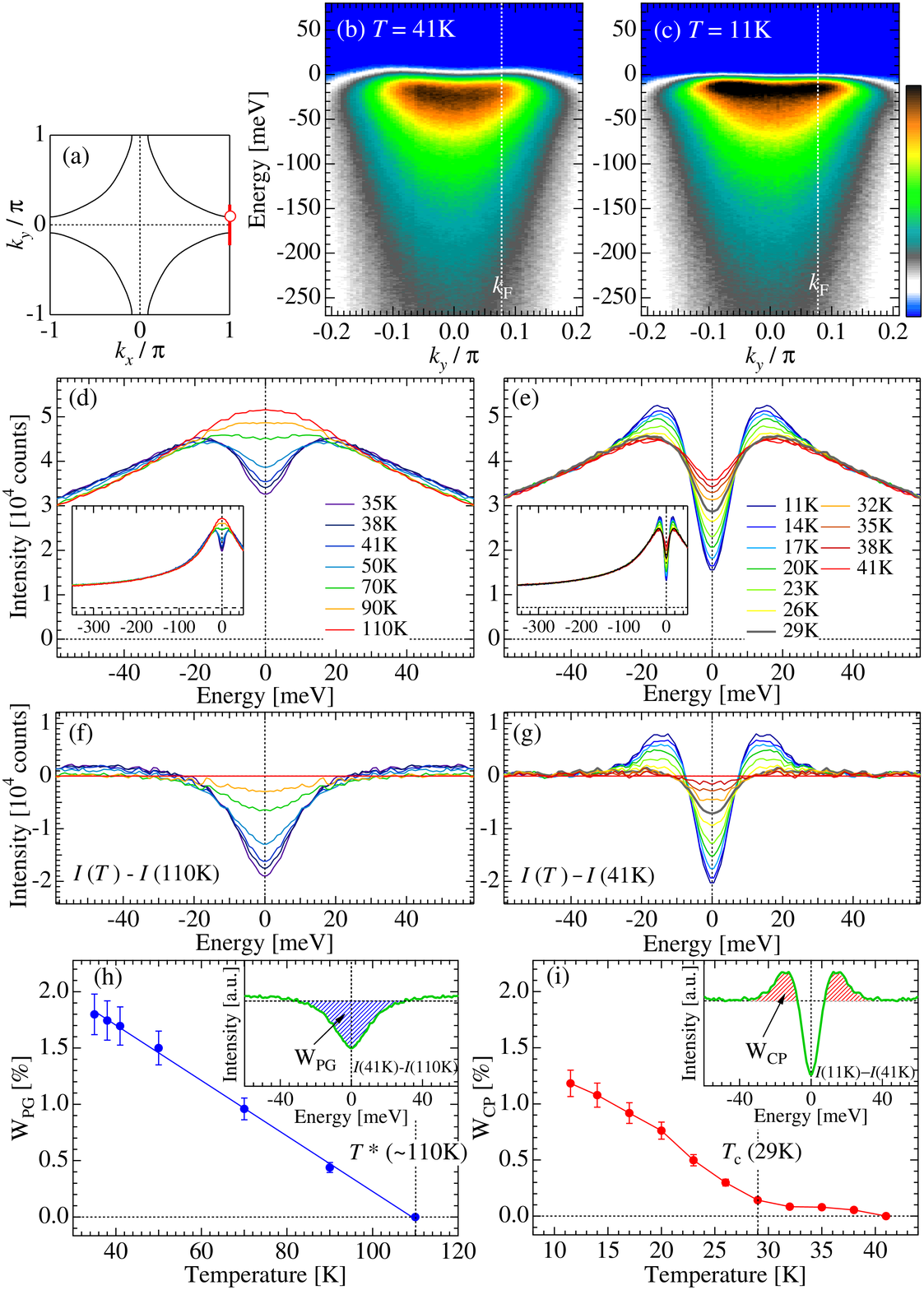}
\caption{{\bf Temperature dependence of the spectral weight in the superconducting and pseudogap states of overdoped Bi2201 ($T_{\rm c}$=29K). Details about the symmetrization and normalization procedures are provided in the Supplementary Information.} {\bf \textsf  a,} Schematic diagram of the Brillouin zone. Red line indicates the cut along which data in panels b-c was acquired. The open circle indicates the antinodal point where data for panels d-e was acquired. {\bf \textsf{b-c,}} ARPES intensity plots along the cut indicated in panel a above and below $T_{\rm c}$ respectively. {\bf \textsf d,} Temperature dependence of the symmetrized EDCs in the pseudogap state for higher temperatures. A single peak at the chemical potential is present at 110K, which corresponds to $T$*. {\bf  \textsf e,} Temperature dependence of the symmetrized EDCs in the supercondcuting state and slightly above $T_{\rm c}$. Note that the coherent peak disappears at $T_{\rm c}$ and spectra in this energy range remain independent of temperature above $T_{\rm c}$. {\bf \textsf f,} Difference spectra obtained by subtracting the EDC at $T$*=110K from the curves in panel d.
{\bf \textsf g,} Difference spectra obtained by subtracting the EDC at 41K from the curves in panel e. {\bf \textsf h,} Temperature dependence of the low energy spectral weight lost in the pseudogap state, the definition of which is shown in the inset.
{\bf \textsf i,} Temperature dependence of the coherent peak weight, the definition of which is
shown in the inset.}
\label{fig1}
\end{figure}
critical temperature and no other significant changes in the spectral
lineshape are observed in this energy range above $T_{\rm c}$. This behavior is
similar to that reported earlier for Bi2212 \cite{Fedorov}. The weight of
the coherent peak has been shown to follow the fraction of the
superconducting electrons or superfluid density \cite{FengScience, HongCoherent} and it is a reasonable measure of the coherence in the system. To conduct a quantitative analysis of the spectral weight we subtract from each EDC in Fig. 1(d) and (e) an EDC obtained at $T^{\ast }$ (110K) or one slightly above $T_{\rm c}$ (41K). The results are shown in Fig. 1(f) and (g). From this data we can easily extract the fraction of the coherent spectral weight and low energy spectral weight lost
due to the pseudogap by defining areas associated with
these features. The selected areas are shown in the insets of Fig. 1(i) and (h).
In Fig. 1(h) we plot the temperature dependence of the spectral weight lost due to the pseudogap opening, as a fraction of total area of the symmetrized EDCs in the range -0.3eV $\leq
E\leq $ 0.3eV (W$_{\rm PG}$($T$)) (see Supplementary Information). We also plot a similar quantity for the coherent spectral weight (W$_{\rm CP}$($T$)) in Fig. 1(i). W$_{\rm PG}$($T$) is linear as a function
of temperature below $T$*, whereas W$_{\rm CP}$($T$) is approximately constant at low temperatures and
zero above $T_{\rm c}$ and only below $T_{\rm c}$ does it rapidly increase with decreasing
temperature. Clearly, the weight of the pseudogap and the coherent peak
behave differently with temperature.

\begin{figure*}
\includegraphics[width=6in]{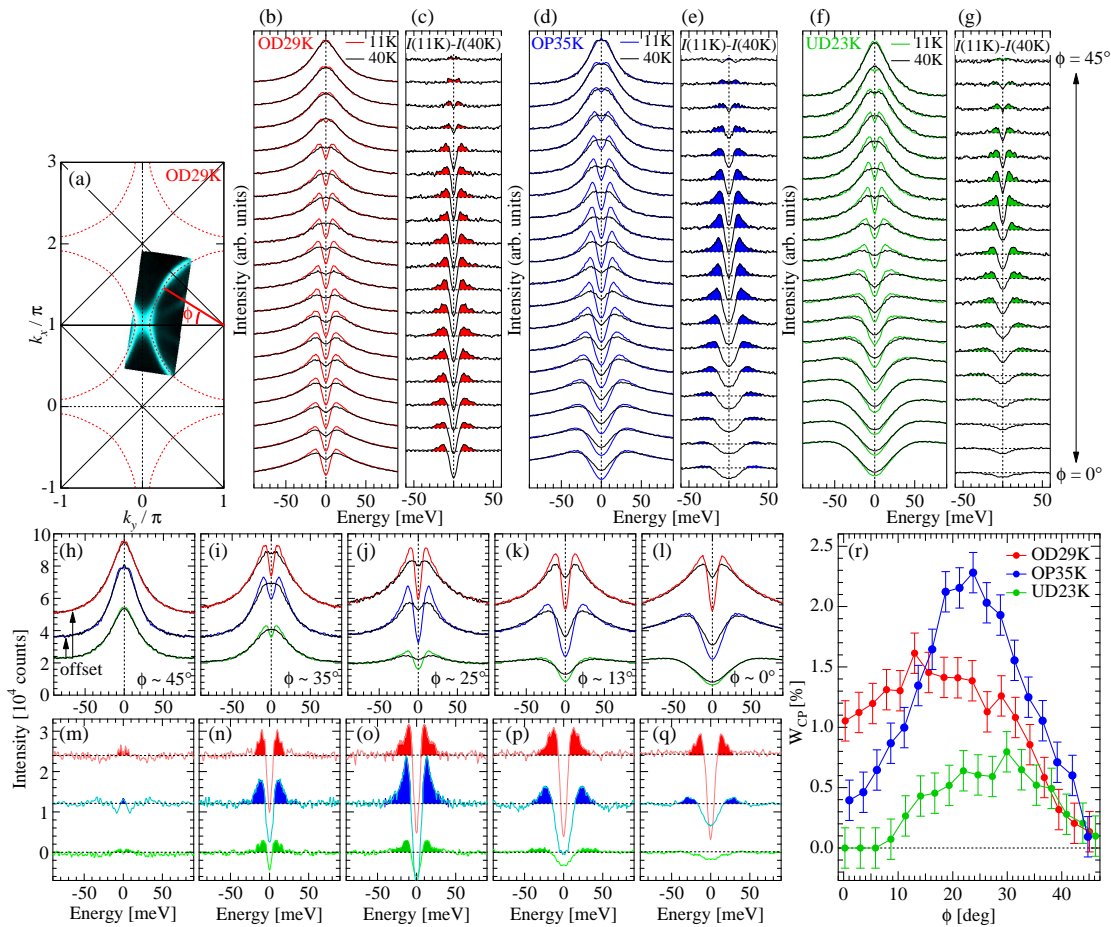}
\caption{{\bf Momentum dependence of the coherent spectral weight
in under-, optimally- and overdoped Bi2201 samples.}
{\bf \textsf{ a,}} Fermi surface map for the OD29K sample and the definition of the Fermi surface angle $\phi$. This plot represents the ARPES
intensity integrated within 10 meV about the chemical potential and measured
at T=40K. The bright areas correspond to the higher photoelectron intensity and
mark the location of the Fermi surface.
{\bf \textsf{b,}} Symmetrized EDCs below and above $T_{\rm c}$ for the overdoped sample ($T_{\rm c}$=29K)
{\bf \textsf{c,}} Difference between the curves in panel b. The area of coherent spectral weight is marked in red.
{\bf \textsf{d,}} Symmetrized EDCs below and above $T_{\rm c}$ for the optimally doped sample ($T_{\rm c}$=35K)
{\bf \textsf{e,}} Difference between the curves in panel d. The area of coherent spectral
weight is marked in blue. {\bf \textsf{ f,}}, Symmetrized EDCs below and above $T_{\rm c}$ for
the underdoped sample ($T_{\rm c}$=23K) {\bf \textsf{ g,}}, Difference between the curves in panel f. The
area of coherent spectral weight is marked in green.
{\bf \textsf{h-l,}} Comparison of the EDCs below (colored curves) and above (black curves) $T_{\rm c}$ for the three doping
levels at a several selected points on the Fermi surface. Arrows in panel h, indicate the offset between the curves used for clarity.
{\bf \textsf{m-q,}} Comparison of the changes in the EDCs across $T_{\rm c}$, obtained from the data in panels h-l. Shaded regions mark the weight of the coherent spectral weight. {\bf \textsf{ r,}}, Momentum
dependence of the coherent spectral weight for the three dopings.}
\label{fig2}
\end{figure*}

We note that by itself the different temperature dependence of W$_{\rm PG}$($T$)
and W$_{\rm CP}$($T$) does not provide information about relation between pseudogap and superconductivity.  We will now use this method to examine the momentum dependence of both quantities. In Fig. 2(b), (d) and (f), we plot the symmetrized EDCs below and above
$T_{\rm c}$ (11K and 40K) measured around the Fermi surface from the node ($\phi =
45^\circ$) to the antinode ($\phi = 0^\circ$) for slightly overdoped
(OD29K), optimally doped (OP35K), and underdoped (UD29K) Bi2201 with $T_{\rm c}$s of
29K, 35K, and 23K, respectively. The evolution of the coherent peak around the Fermi surface (Fig. 2(a)) can be visualized by plotting the difference between the EDCs below and above $T_{\rm c}$ as shown in the Fig. 2(c), (e), and (g). The shaded regions in these difference curves mark the area of the coherent spectral weight (W$_{\rm CP}$, see the inset of Fig. 1(h)). Remarkably, the weight of the coherent peak has a very unexpected momentum dependence, which varies significantly with doping. In the overdoped sample (Fig. 2(c)), the
weight of the coherent peak increases away from the node (top curve) and it
saturates near the antinode (bottom curve). In contrast, W$_{\rm CP}$ for the
optimally and underdoped samples (Fig. 1(e) and (g), respectively) is highly
non-monotonic. It initially increases near the node, just as in the overdoped
case, but then it is abruptly suppressed in the antinodal region. This is
quite remarkable, since for all our samples the magnitude of the superconducting
gap follows a $d$-wave symmetry consistent with other reports\cite{ZhouBi2201, Ming}, so one might expect the weight of the coherent peak to remain constant or increase monotonically around the Fermi surface with the largest value at the antinode for all dopings. To better visualize the doping dependence of this unusual behavior, we plot the symmetrized EDC data and difference curves for the three doping levels in Fig. 2(h)-(q). Again the areas of the coherent spectral weights are shaded in the difference spectra (Fig. 2(m)-(q)). We quantize these results by plotting W$_{\rm CP}$ in Fig. 2(r) for the three samples as a function Fermi angle $\phi$. We note that the coherent peak is present at antinode in optimally and overdoped samples consistent with previous results \cite{ZhouBi2201, FengBi2201}. Our careful measurements reveal that suppression of the coherent peak near the antinode ($\phi$=0) occurs even in the overdoped samples and becomes stronger with underdoping. 

\begin{figure}
\includegraphics[width=3.6in]{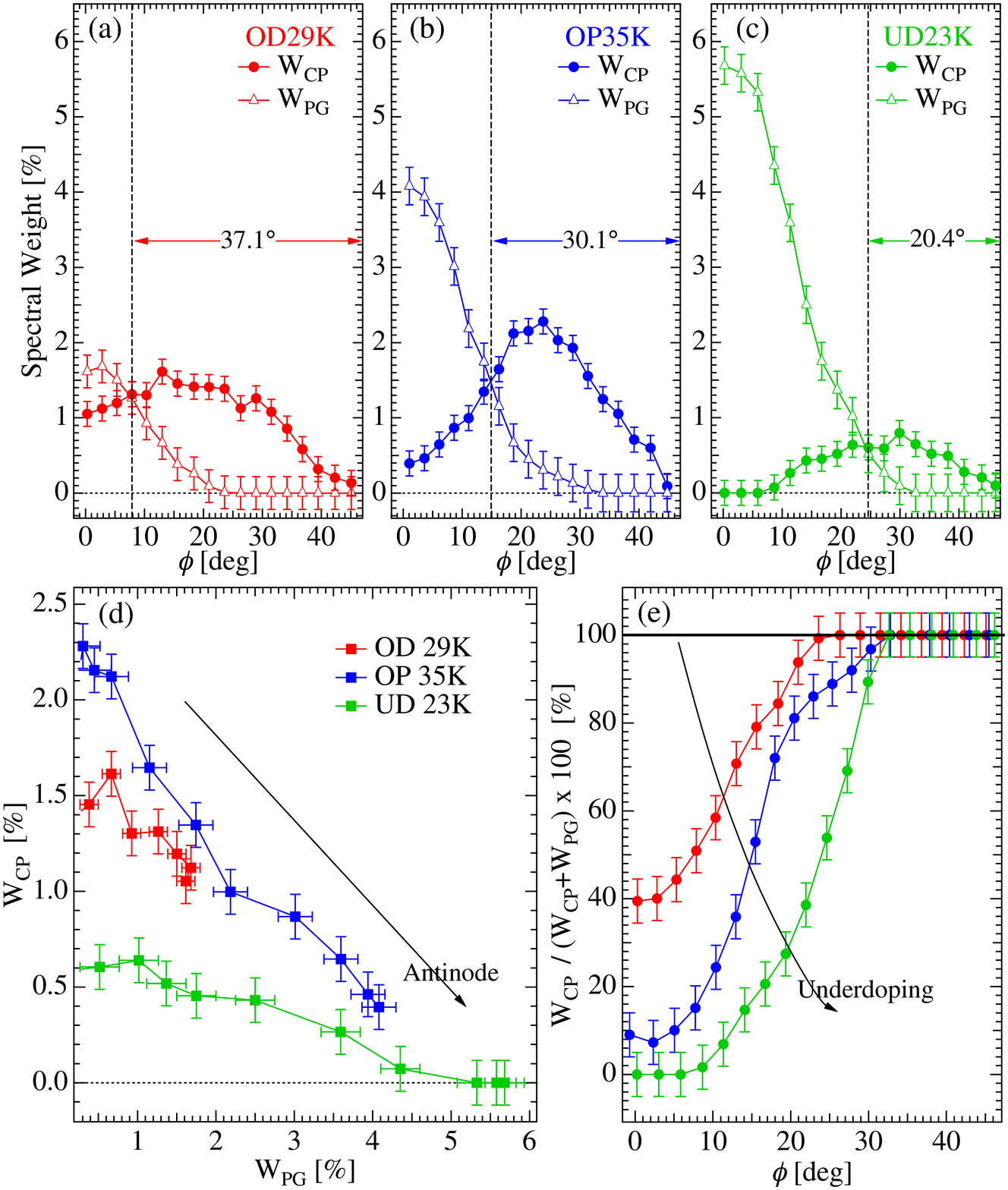}
\caption{{\bf Momentum dependence of the coherent and pseugodap
spectral weight in under-, optimally- and overdoped Bi2201 samples.}
{\bf \textsf{a-c,}} Spectral weight lost due to the pseudogap (open triangles) and the coherent
spectral weight (filled circles) plotted as a function of the Fermi surface
angle expressed as a percentage of the total spectral weight integrated within
$\pm$300 meV from chemical potential for the overdoped ($T_{\rm c}$=29K), optimally doped ($T_{\rm c}$=35K) and underdoped ($T_{\rm c}$=23K) samples, respectively. The arrows indicate the FS angle range of the coherent Fermi surface (i. e. where the coherent peak weight dominates over the pseudogap weight). Note that this range shrinks with underdoping.
{\bf \textsf{d,}} Plot of the coherent spectral weight vs. spectral
weight lost due to the pseudodap. {\bf \textsf{e,}} Ratio of the coherent spectral weight to the total change of the spectral weight for the
three doping levels.}
\label{fig3}
\end{figure}

\begin{figure}
\includegraphics[width=3.1in]{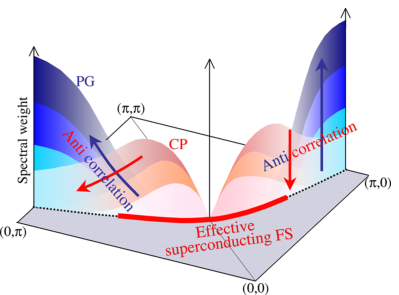}
\caption{{\bf Schematic illustration of the momentum and doping evolution of
the coherent and pseudogap spectral weight and the effective region of the
superconducting quasiparticles.} The arrows mark the anticorrelation between pseudogap and high temperature superconductivity in momentum space and with varying doping level.}
\label{fig4}
\end{figure}

In Fig. 3(a), (b) and (c) we extract the spectral weight lost near the
Fermi level due to the pseudogap opening (W$_{\rm PG}$, see the inset of Fig. 1(h))
and compare its momentum dependence with that of the coherent spectral
weight (W$_{\rm CP}$) for the three dopings.
W$_{\rm PG}$ is zero near the node for all three dopings because the
superconducting gap closes at $T_{\rm c}$, creating a Fermi arc. W$_{\rm PG}$ increases
towards the antinode and reaches a maximum there. This behavior contrasts
with that of the coherent peak weight. In fact, the onset of the suppression
of the coherent weight coincides with the increase of the weight lost to the
pseudogap, consistently for all dopings. To examine the relationship between
the two quantities in more detail, we plot W$_{\rm CP}$ vs W$_{\rm PG}$ in Fig. 3(d).
Surprisingly, we find an almost perfect linear anti-correlation between the
two quantities in all three samples. A similar anti-correlation can be also
demonstrated as a function of doping. In Fig. 3(e) we plot the ratio of the
coherent spectral weight to the total change of the spectral weight (W$_{\rm CP}$+W$_{\rm PG}$). Here the angle $\phi$ where the pseudogap dominates the spectral
lineshape and suppresses the coherent component becomes smaller (closer to
the node) as the doping decreases, and both the pseudogap and $T$* become
larger. This is clear evidence that the two orders: the pseudogap and the
superconducting coherence compete for the low energy spectral weight. This
also means that the coherent part of the Fermi surface in the
superconducting state shrinks with decreasing carrier concentration.

In Fig. 4, we schematically sketch the anti-correlation between the pseudogap and coherent peak in momentum space. The competition between the two leads to a highly unusual superconducting state in underdoped regime, where only part of the Fermi surface displays any sign of coherence. We note that pseudogap formation due to preformed Cooper pairs is clearly inconsistent with our results. In that case, a spectrum with a pseudogap above $T_{\rm c}$ should transform into that of a coherent superconductor below $T_{\rm c}$, contrary to our findings. Similarly, in the resonating valence bond picture \cite{Anderson}, at least in its most elementary version, the superconducting coherency should be positively correlated with the pseudogap behavior. Instead, our data are consistent with the view that the pseudogap is due to the formation of a ordered dimer state\cite{Sachdev} or a density wave state\cite{DavisCDW, EricCDW} that partially gaps the Fermi surface. Then superconducting coherency below $T_{\rm c}$ can only emerge in the remaining parts of the Fermi surface. Such a density wave could exist homogeneously or, as numerous  STM/STS experiments suggest, emerge in an inhomogeneous fashion\cite{Davis}. 
In the latter case, our results put strong constraints on the nature of the inhomogeneous state: they are inconsistent with the scenario where underdoped and overdoped regions of ordinary $d$-wave supercondutor coexists in real space with different superconducting gap amplitude. In such a situation the weight of the coherent peak would not display the observed non-monotonic momentum dependence. The same observation also excludes the view that independent pseudogap and superconducting regions exist and requires a penetration of the superconductivity into the pseudogapped regions. In essence, the pseudogap competes with the high temperature superconductivity not only in real space but also in the momentum space.

A pseudogap that competes with the superconductivity has implications for the
anomalous doping and temperature dependence of the superfluid stiffness $
\rho _{\rm s}\left( x,T\right) $ of the cuprates\cite{Lee, Ioffe}. Penetration depth
experiments observe a strong doping dependence of $\rho _{\rm s}\left(
x,T=0\right) $, while the low temperature  slope $\left. d\rho
_{\rm s}/dT\right\vert _{T=0}$ \ is weakly doping dependent \cite{Lemberger}. The latter
implies that the contribution to $\rho _{\rm s}$ from states close to the node
is weakly doping dependent. Our results offer an explanation for the
reduction of $\rho \left( x,T=0\right) $ as due to the suppression of coherency at the
antinodes.  $\rho _{\rm s}\left( x,T=0\right) $ should  be roughly proportional
to the angle range $45^{o}-\varphi _{\max }\left( x\right) $ beyond which
the coherent weight is suppressed, a prediction that can be easily verified.

\bigskip

{\bf Supplementary Information} is linked to the online version of the paper at www.nature.com

\bigskip

{\bf Acknowledgements} We thank Andrew J. Millis and Helen M. Fretwell for useful discussions . This work was supported by Basic Energy Sciences, US DOE. The Ames Laboratory is operated for the US DOE by Iowa State University under Contract No. W-7405-ENG-82.

\bigskip

{\bf Authors contributions}
T. K., R. K. and A. K. developed concept of the experiment. T. K. and T. T. grew high quality single crystals. T. K. acquired experimental data, T. K. and A. K. performed data analysis. T. K., A. K. and J. S. wrote the manuscript.

{\bf Author Information}
Reprints and permissions information is available at www.nature.com/reprints. Correspondence and requeests for materials should be addressed to A. K. (kaminski@ameslab.gov) or T. K. (kondo@ameslab.gov)

\end{document}


\title{Supplementary Information: Direct evidence for a competition between the pseudogap and high temperature superconductivity in the cuprates}

\author{Takeshi Kondo}
\affiliation{Ames Laboratory and Department of Physics and Astronomy, Iowa State University, Ames, IA 50011, USA}

\author{Rustem Khasanov}
 \affiliation{Laboratory for Muon Spin Spectroscopy, Paul Scherrer
Institut, CH-5232 Villigen PSI, Switzerland}

\author{Tsunehiro Takeuchi}
\affiliation{Department of Crystalline Materials Scherrerience, Nagoya University,
             Nagoya 464-8603, Japan}
\affiliation{EcoTopia Science Institute, Nagoya University, Nagoya 464-8603, Japan}

\author{Joerg Schmalian}
\affiliation{Ames Laboratory and Department of Physics and Astronomy, Iowa State University, Ames, IA 50011, USA}

\author{Adam Kaminski}
\affiliation{Ames Laboratory and Department of Physics and Astronomy, Iowa State University, Ames, IA 50011, USA}

\date{\today}

\maketitle

{\bf Samples and Experimental method of ARPES}

(Bi,Pb)$_2$(Sr,La)$_2$CuO$_{6+\delta}$ (Bi2201) single crystals
were grown by the conventional floating-zone (FZ) technique.
Bi2201 is a single layered cuprate (i.e. it has a single CuO$_2$ plane per unit cell), therefore its spectra are free from bilayer-induced band splitting.
We partially substituted Pb for Bi to suppress the modulation in the BiO plane, which causes a contamination of the ARPES signal. The outgoing photoelectrons are diffracted, creating multiple images of the band and Fermi surface that are shifted in momentum. The bilayer-free and modulation-free samples enable us to precisely analyze the ARPES spectra. We controlled the carrier concentration of the over-, optimally-, and underdoped samples with $T_{\rm c}$=29K, 35K, and 23K (OD29K, OP35K, and UD23K), respectively, by the partial substitution of La for Sr and subsequent annealing.
The nominal composition and annealing condition of those three samples are summarized in Table I.

\begin{figure}
\includegraphics[width=2.9in]{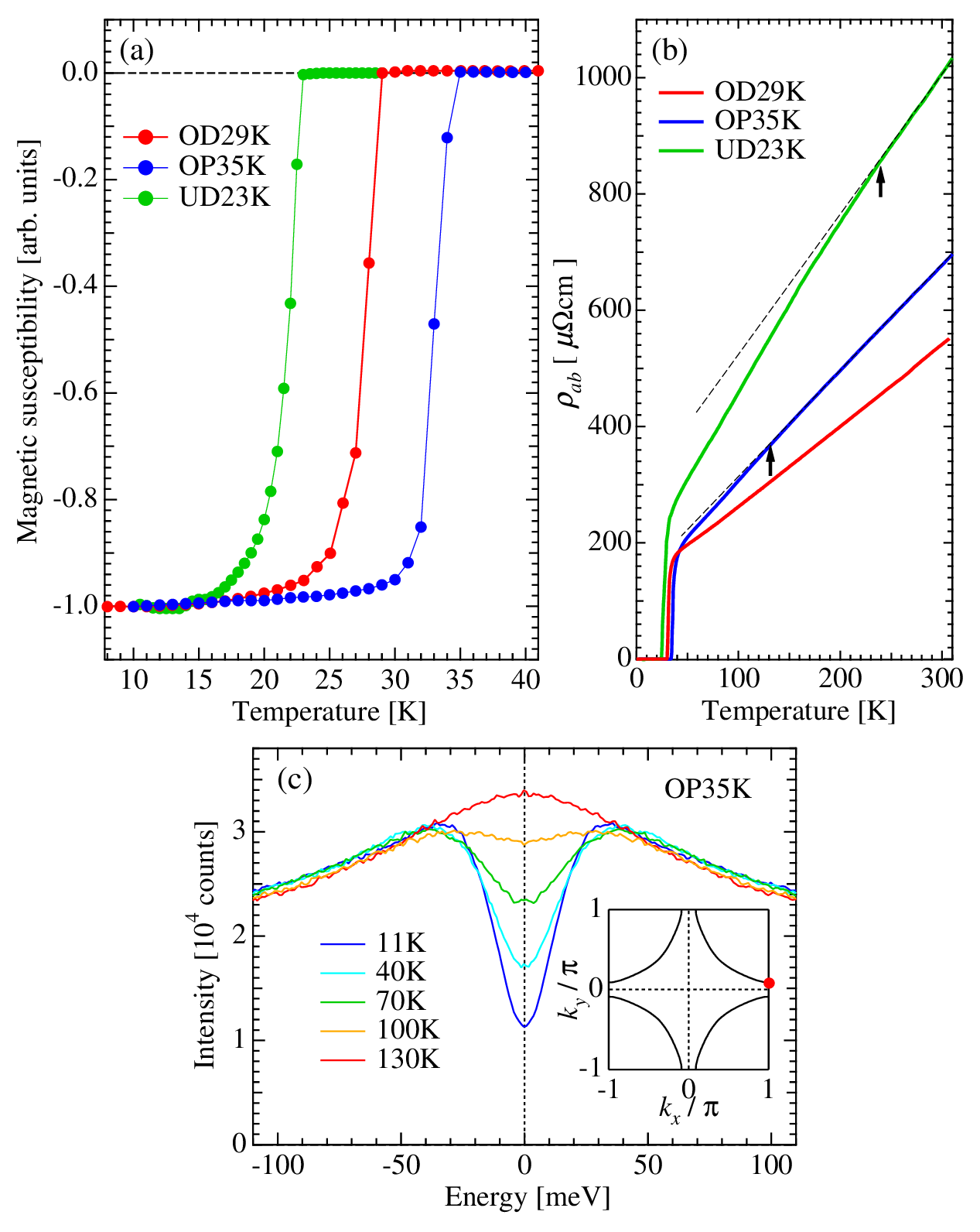}
\caption{Characterization of the samples using transport and susceptibility measuremnts. 
{\bf \textsf{a,}} Magnetic susceptibility for OD29K, OP35K and UD23K samples. 
{\bf \textsf{b,}} Temperature dependence of resistivity for OD29K, OP35K and UD23K samples.  Dashed lines in panel b are fitted to the resistivity at high temperatures for OP35K and UD23K. The deviation from the $T$-linear behavior (marking the onset of the pseudogap) is indicated with arrows. 
{\bf \textsf{c,}} Symmetrized EDCs of OP35K measured at the antinodal Fermi momentum (marked with a circle in the inset) for temperatures ranging from deep below $T_{\rm c}$ (11K) to the temperature where the pseudogap closes ($T$*$\sim$130K).}
\label{figS1}
\end{figure}

\begin{figure}
\includegraphics[width=2.7in]{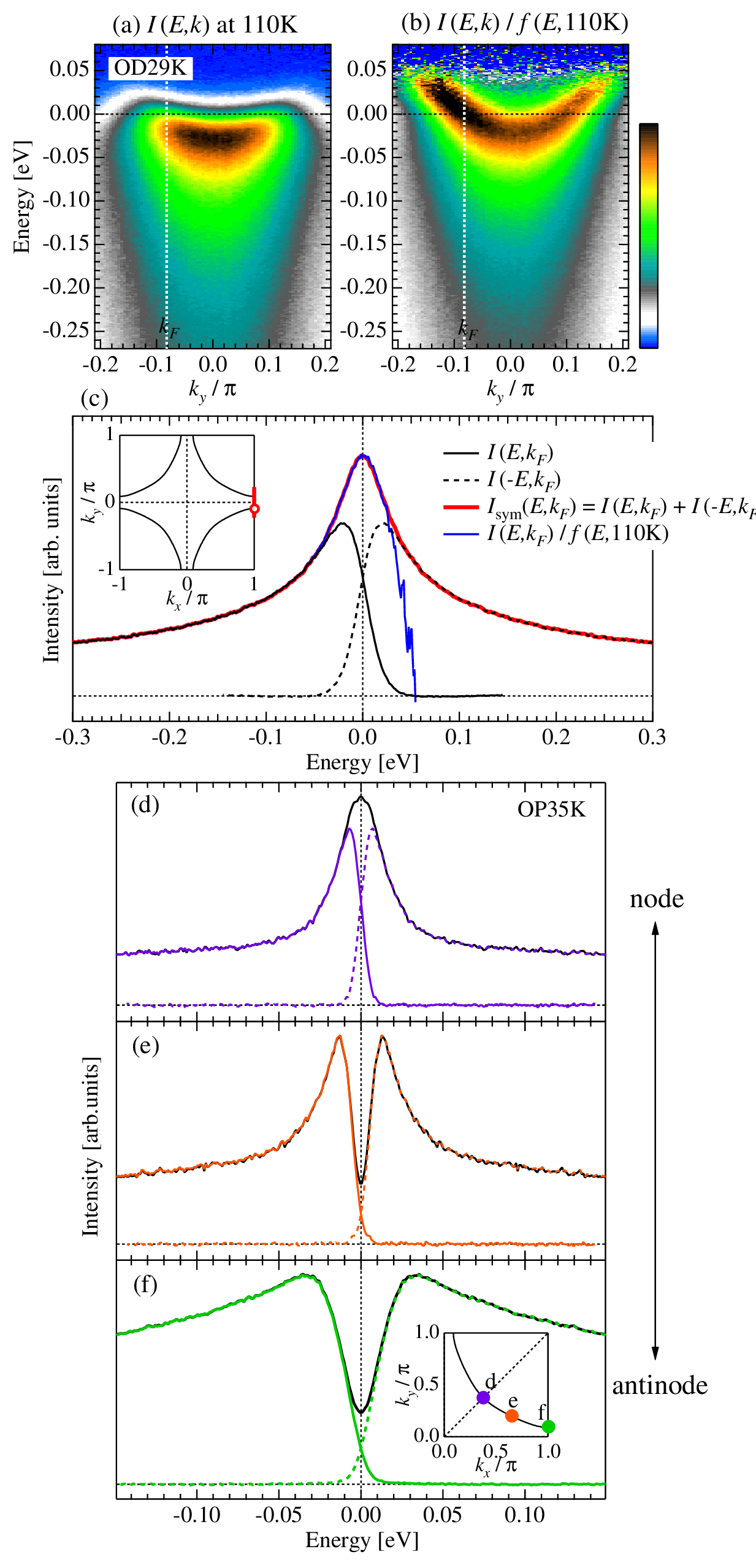}
\caption{Demonstration of symmetrization method.
{\bf \textsf{a,}} ARPES intensity map as a function of energy and $k_y$, $I(E,k)$, measured at $T$=110K along a momentum cut drawn in the inset of c. {\bf \textsf{b,}} ARPES intensity map from a, divided by Fermi function, $I(E,k)/f(E, $110K). {\bf \textsf{c,}} EDC (black curve), EDC reflected about the chemical potential (dashed black curve), and symmetrized EDC (red curve, sum of black and dashed black curves) are compared to EDC divided by Fermi function (blue curve). All curves were obtained at Fermi momenutm marked by a circle in the inset. The plain EDC and EDC divided by the Fermi function correspond to intensity along the dashed white line in panel a and b, respectively. The curves obtained from the symmetrization and division by the Fermi function methods agree very well up to an energy comparable to 2 $k_{\rm B}T$ ($\sim$20 meV).}
\label{figS2}
\end{figure}

\begin{figure}
\includegraphics[width=2.6in]{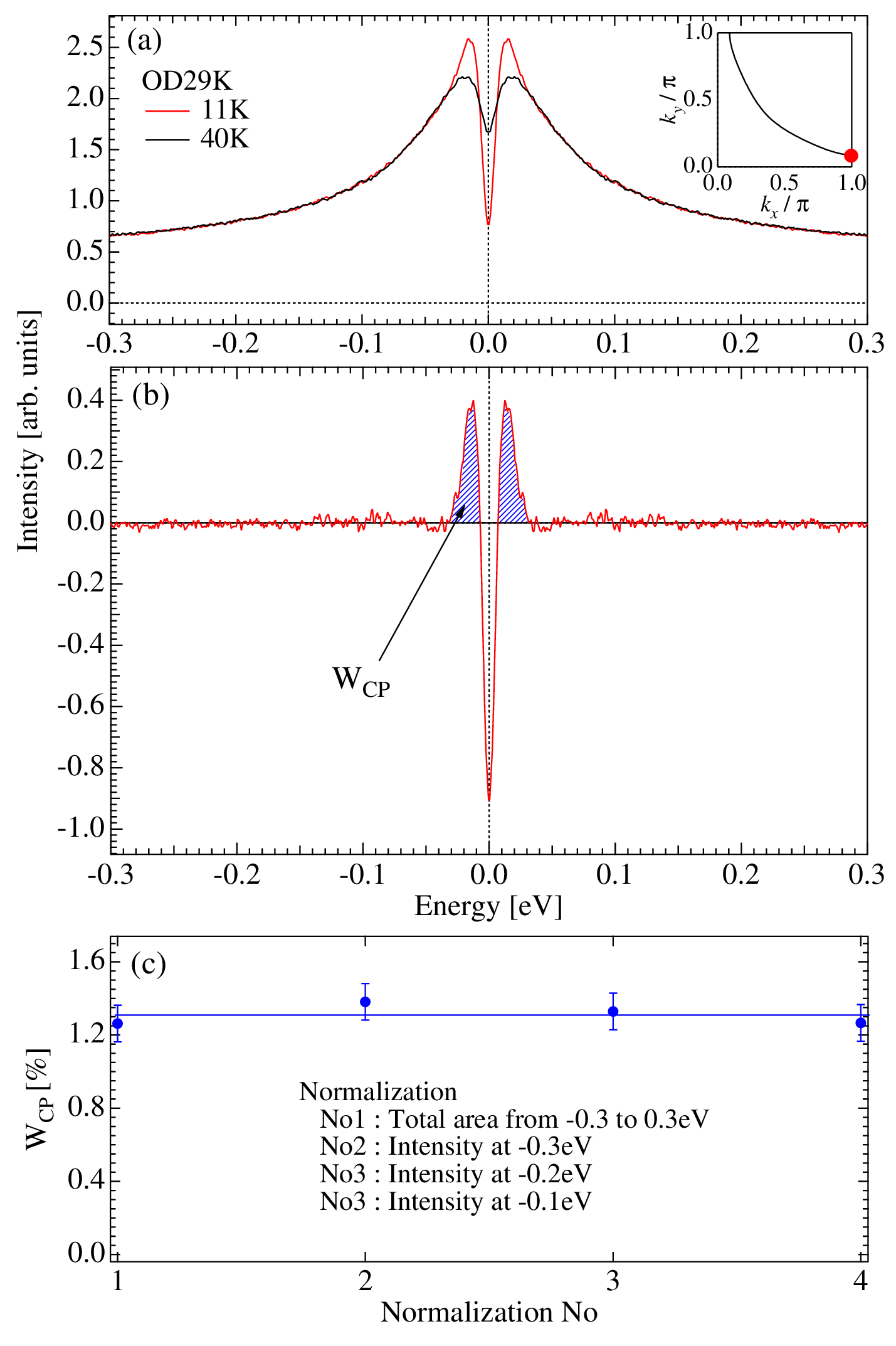}
\caption{Verification of the normalization procedure. {\bf \textsf{a,}} Symmetrized EDCs of OD29K deep below $T_{\rm c}$ (11K) and above $T_{\rm c}$ (41K) measured at Fermi momentum marked in the inset. Both spectra are normalized to a whole area from -0.3eV to 0.3eV. {\bf \textsf{b,}} Subtracted spectrum between two symmetrized EDCs shown in a. The blue hatched area corresponds to a coherent spectral weight (W$_{\rm PC}$). {\bf \textsf{c,}} W$_{\rm PC}$ shown as a percentage to the total area of symmetrized EDC using four different normalization methods. Normalization No1 uses the total area integrated from -0.3eV to 0.3eV. No2, No3 and No4 use area integrated within $\pm$ 30 meV centered at -0.3eV, -0.2eV and -0.1eV, respectively. Results are within the error bars used in this letter}
\label{figS3}
\end{figure}

Figure 1(a) and (b) shows the magnetic susceptibility and resistivity of OD29K ($T_{\rm c}$=29K), OP35K($T_{\rm c}$=35K), and UD23K. All samples show a sharp superconducting transition ($\sim$5K).
The signature of the pseudogap is observed clearly in the resistivity of OP35K, and UD23K: $T$-linear behavior observed at high temperature changes its slope at $T$* $\sim 130$K, and $\sim 240$K for OP35K and UD23K, respectively. This is less clear for the OD29K sample because of the proximity of $T$*  and $T_{\rm c}$. Fig. 1(c) shows the symmetrized EDCs for OP35K measured at the antinodal Fermi crossing point (circle in the inset of panel (c)) from deep below $T_{\rm c}$ (11K) to the pseudogap closing temperature ($\sim 130$K).  We find that the particular temperature of $T$*  observed in the resistivity measurement directly corresponds to the pseudogap closing temperature.  One drawback of studying the pseudogap in high $T_{\rm c}$ cuprates such as Bi2212 is the large energy and temperature scale of the superconducting gap ($\sim$40 meV and $\sim$90K, respectively, at optimal doping), which is comparable to those of the pseudogap.
This similarity makes it difficult to separately investigate these two phenomena.
We chose to study Bi2201, which has a low $T_c$ of $\sim$35K and a large $T$*  (similar to that of Bi2212 at optimal doping) in order to gain an important insight into the relationship between the pseudogap state and the superconductivity.

\begin{table*}
\caption{Sample preparation conditions for the present Bi2201 single crystals.}
\begin{center}
   \begin{tabular}{c|c|c|c|c|c}
    \hline
    \hline
     label & $T_c$(K) & nominal composition & atmosphere & temperature\\
    \hline
     OD29K & 29 & (Bi$_{1.35}$Pb$_{0.85}$)(Sr$_{1.47}$La$_{0.38}$)CuO$_{6+\delta}$ & Air & 450C$^\circ $ for 24hours\\
     OP35K & 35 & (Bi$_{1.35}$Pb$_{0.85}$)(Sr$_{1.47}$La$_{0.38}$)CuO$_{6+\delta}$ & Ar flow & 650C$^\circ $ for 24hours\\
     UD23K & 23 & (Bi$_{1.1}$Pb$_{1.0}$)(Sr$_{1.28}$La$_{0.6}$)CuO$_{6+\delta}$ & vacuum & 650C$^\circ $ for 24hours\\
    \hline
    \hline
\end{tabular}
\end{center}
\end{table*}

The ARPES data was acquired using a laboratory-based system consisting of a Scienta SES2002 electron analyzer and GammaData helium UV lamp. All data were acquired using the He I line with a photon energy of 21.2 eV. The angular resolution was $0.13^\circ$ and $\sim 0.5^\circ$ along and perpendicular to the direction of the analyzer slits, respectively. The energy resolution was set at $\sim5$meV. The energy corresponding to the chemical potential was determined from the Fermi edge of polycrystalline Au in electrical contact with the sample. The stability of the Fermi edge was better than 0.3 meV during all measurements. Custom designed refocusing optics enabled us to accumulate high statistics spectra in a short time without being affected by possible sample surface aging due to the absorption or loss of oxygen. Special care was taken in the purification of the helium gas used in the UV source to remove even the slightest amount of contamination that could contribute to surface contamination of the sample. Typically no changes in the spectral lineshape were observed in consecutive measurements over several days. The samples were cooled using a closed-cycle refrigerator. Measurements were performed on several samples and we confirmed that all yielded consistent results.

{\bf Symmetrization method}

Normally ARPES spectra are cut off by the Fermi function near the chemical potential. It is important to remove the temperature dependent Fermi cut-off from the spectra in order to investigating intrinsic temperature dependence of the spectral function. The most straightforward way to achieve this is to divide the ARPES spectra by the Fermi function. Figure 2(b) plots a result of this treatment for the ARPES intensity map shown in  Fig. 2(a), which was measured for OD29K along a momentum cut represented in the inset of panel (c). The unoccupied band dispersion clearly shows up after dividing by the Fermi function at a temperature of 110K. However, this method works only well at high temperatures, where there is significant spectral weight beyond the chemical potential.  At low temperatures, the Fermi function sharply drops to nearly zero beyond chemical potential, and thus division of the ARPES spectra by very small values amplifies the noise and renders that part of the extracted spectral function useless. Since most of data presented here focuses on low temperatures, we used the alternative symmetrization method proposed by Norman \cite{NormanNature}. This method is not only valid for the high temperature data, but also it is effective for removing the Fermi cut-off from the low temperature data.
In Fig. 2(c), we compare results from the two methods (dividing and symmetrization) using the EDC for OD29K measured at the Fermi closing point (red circle in the inset of panel (c)). Perfect agreement is seen near the Fermi level, which indicates that
the symmetrization method successfully removes the Fermi cut off and, in the case of our samples, does not suffer from problems related to particle-hole asymmetry.
Note that the result of the dividing method is not useful above $\sim$0.03eV due to the noise - even for very high statistics data.
Panel (d), (e), and (f) demonstrates the process of symmetrization using EDCs of OP35K measured at three different Fermi crossing points from the node to the antinode (dots in the inset of panel (f)). A single peak is observed at the node indicating that there is no energy gap. Away from the node, the spectra have two peaks centered at the Fermi level due to a gap opening. We extracted a coherent spectral weight from the symmetrized EDC with two peaks to take into account the existence of BCS-like Bogoliubov Quasiparticles in the cuprates \cite{Campuzano,Matsui}, which at the Fermi momentum have two coherent peaks, symmetric about the chemical potential with the same spectral weights.

{\bf Normalization of ARPES spectra}

We express the weight of both the low energy spectral weight lost to the pseudogap and the weight of the superconducting coherent peak, as a fraction of total spectral weight. Such an approach ensures that our results are not affected by matrix elements which normally cause a variation in the spectral intensity with momentum. However, when extracting the weight of either quantity, we need to subtract two EDCs obtained at different temperatures. Using the raw data divided by the acquisition time and photon flux seems to be the best approach, however it produces some noise because the photon flux varies slightly in time. For the purpose of the discussion in this letter, we normalized all the symmetrized EDCs to the total area of each spectrum in the range -0.3eV $\leq E\leq $ 0.3eV. Since we are only interested in the behavior of the very low energy spectral weight with temperature, it is reasonable to assume that the spectral intensity does not change significantly at energies much higher than temperature energy scale. This is demonstrated in Fig. 3(a), where we plot two EDCs obtained at 11K and 40K from the OD29K sample. The spectral lineshape is identical in both cases for energies larger than $\sim$30 meV. This is demonstrated also in Fig. 3(b), where the difference between two EDSs is zero within the noise beyond $\sim$30 meV. We have also verified explicitly that the particular normalization scheme does not affect the extraction of the low energy fraction of the spectral weight. 
We compared the result of total area normalization with those of three narrow energy range normalizations, which were 
 performed within $\pm$30 meV centered at -0.3eV (No2),  -0.2eV (No3), and -0.1eV (No4).
The weight of the superconducting coherent peak extracted after those three normalizations is plotted in the panel (c) along with that obtained for the total area normalization. It is clear that results corresponding to all four normalization schemes fall well within the error bars. These results prove that the value of coherent spectral weight (W$_{\rm CP}$) does not significantly depend on how one normalizes the ARPES spectra. All different normalization methods yield similar results simply because the the spectra do not vary significantly with temperature at higher energies as shown in Fig. 3(c).